\documentclass[twocolumn,journal]{IEEEtran}

\usepackage{graphicx}
\usepackage{amsmath}
\usepackage{amssymb}
\usepackage{comment}

\usepackage{amsthm}
\usepackage{cite}
\usepackage{float}
\usepackage{color}
\begin{document}

\title{\LARGE A Secure Multiple-Access Scheme for Rechargeable Wireless Sensors in the Presence of an Eavesdropper}
\author{Ahmed El Shafie and Naofal Al-Dhahir, \emph{Fellow, IEEE}\\
\begin{tabular}{c}
 {University of Texas at Dallas, USA}
\thanks{This paper was made possible by NPRP grant number 6-149-2-058 from the Qatar National Research Fund (a member of Qatar Foundation). The statements made herein are solely the responsibility of the authors.}
\thanks{This paper is published in IEEE Communications Letters 2015.}
\end{tabular}}
\date{}
\maketitle
\begin{abstract}
We propose a simple yet efficient scheme for a set of energy-harvesting sensors to establish secure communication with a common destination (a master node). An eavesdropper attempts to decode the data sent from the sensors to their common destination. We assume a single modulation scheme that can be implemented efficiently for energy-limited applications. We design a multiple-access scheme for the sensors under secrecy and limited-energy constraints. In a given time slot, each energy-harvesting sensor chooses between sending its packet or remaining idle. The destination assigns a set of data time slots to each sensor. The optimization problem is formulated to maximize the secrecy sum-throughput.
\end{abstract}
\begin{IEEEkeywords}
Energy harvesting, fading channels, eavesdropper, secrecy sum-throughput.
\end{IEEEkeywords}
\vspace{-0.3cm}
\section{Introduction}
\vspace{-0.1cm}
The energy harvesting capability enables battery charging by ambient energy sources instead of frequent battery replacements.
An energy-harvesting sensor (EHS) is a promising solution for
maintenance-free wireless sensor networks. It relies only on the energy harvested from ambient energy sources such as solar, vibration, thermal, and radio frequency\cite{5522465}
for its operation. The design of an EHS presents
new challenges compared to a battery-based sensor because the
energy availability from the environment is likely to be sporadic and random \cite{reddy,5522465}.

The authors of \cite{ref8} and \cite{ref10} consider both the energy harvesting
profile and the channel statistics in scheduling EHS transmissions. Both works focus on
energy allocation for EHS transmissions assuming that the energy level and channel state change randomly from one slot to another. The
proposed approaches in \cite{ref8} and \cite{ref10} allow
the EHS to transmit at variable data rates depending
on the battery energy level and channel state. We assume a single
modulation scheme \cite{reddy}, which is a typical restriction in
communication standards designed for energy-efficient applications \cite{reddy}. This allows the EHS hardware to
be highly optimized from an energy consumption perspective.
Using adaptive modulation schemes as in \cite{ref8,ref10,6657835}
would improve the average achieved EHS data rate but at the
cost of increased implementation complexity.

Security issues in wireless communication
networks have received significant attention recently \cite{4035982}. Unlike \cite{reddy,ref8,ref10,6657835}, we consider secure communication among energy-constrained sensors and design multiple-access schemes to manage the access of the different network's sensors. In addition, we take into account the impact of interference from other sensor nodes that use the same unlicensed frequency band.

In this letter, we assume a set of sensors that communicate with a common destination in the presence of an eavesdropper. Time and channel access are slotted. The eavesdropper attempts to decode the sensors' messages in each time slot. Our contributions in this letter are as follows: a) Unlike \cite{ref8,ref10,6657835}, we consider the multi-sensor scenario and take the security issue into consideration in addition to the processing energy consumed at the EHS and the interference at each sensor due to other sensors which use the same unlicensed frequency band; b) we propose an
information-theoretic formulation of the problem of secure communication
between the sensors and the destination; c) we propose a new multiple-access scheme for fixed-rate systems where sensor nodes only send packetized data using a single modulation scheme, which is common for energy-constrained applications \cite{reddy}. Our proposed scheme allocates sensors to data time slots based on security considerations as well as battery energy levels during each frame; d) we derive the maximum secure sum-throughput of the sensor network.
\vspace{-0.2cm}
         \section{System Model and Assumptions}
\begin{figure}
  \centering
  \includegraphics[width=0.7\columnwidth]{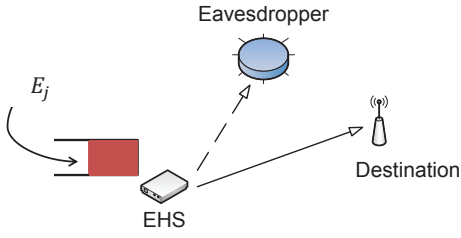}\\
  \caption{Network model with one sensor, the eavesdropper and the destination. $E_j$ denotes the energy harvested by the sensor in time slot $j$.}
  \label{fig00}
  \vspace{-0.6cm}%
  \end{figure}

         \vspace{-0.15cm}
We consider a network composed of $\mathcal{M}$ sensors communicating with a single destination. The sensor transmissions use the unlicensed industrial, scientific and medical (ISM) frequency bands. Each frame is divided into three time intervals: (1) a time interval for channel estimation and information exchange between the sensors and the destination; (2) a time interval for transmitting the beacon signal which contains the assignment (allocation) of data time slots to sensors during the frame from the destination to the sensors; and (3) a time interval for the data time slots. The destination sends a beacon that indicates the beginning of the data time slots and shows the distribution of time slots among sensors. It is assumed that the activities during communication between the sensors and destination deplete a certain amount of energy, denoted by $\tilde E_\tau$ energy units for frame $\tau$. We assume that if the communication for a sensor is not secured, the destination will not assign any time slots for that sensor. Each frame lasts for $T_f$ seconds. The length of the data time slots is $T<T_f$ slots each of length {\it one second}. We refer to the sensors as the legitimate sources. Each sensor communicates with the legitimate destination (master node) in the presence of an eavesdropper as shown in Fig. \ref{fig00}.

\begin{figure}
  \centering
  \includegraphics[width=0.89\columnwidth]{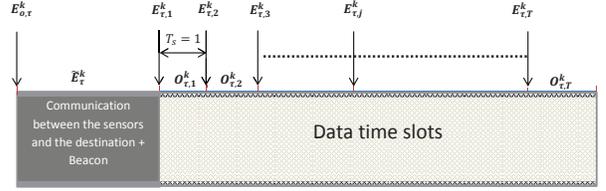}\\
  \caption{Slotted system model. The figure shows one frame. The energy harvested in time slot $j$ within frame $\tau$ is denoted by $E^k_{\tau,j}$, whereas the energy transmitted in time slot $j$ within frame $\tau$ is denoted by $o^k_{\tau,j}$.}
  \label{fig0}
  \vspace{-0.5cm}%
  \end{figure}

The sources are rechargeable and capable of harvesting energy from the ambient energy sources. The energy harvested by a sensor is stored (buffered) in a rechargeable battery (energy queue). At times $t=1,2,\dots,T,$ the $k$th sensor collects energy
with amounts $E^k_{\tau,1},E^k_{\tau,2},\dots,E^k_{\tau,T}$ \cite{6657835}. That is, the cumulative energy harvested at Sensor $k$ from the ambient energy sources in time slot $j$ of frame $\tau$ is $E^k_{\tau,j}$ energy units (see Fig.~\ref{fig0}). The sources are saturated, i.e., always backlogged with data packets. In addition, data and energy buffers are assumed to be unlimited in size \cite{6657835}.

Unlike \cite{ref8} and \cite{6657835}, we assume that energy is consumed due to data transmissions as well as any other energy costs such as processing, measurements, sensing, circuitry, etc, whose cumulative energy is assumed to be $E^k_{\rm p}$ energy units per data time slot. Moreover, we assume that the energy amounts are known a priori which is justified if the environment is highly predictable, e.g., the energy is harvested from a solar panel or from the vibration of motors that are turned on only during fixed operating hours \cite{6657835}. To conserve the sensor energy, the sensor is ON in a given time slot only when it is assigned to that slot.

We assume that each sensor is assigned a set of time slots within each frame. That is, during frame $\tau$, Sensor $k$ is assigned $\omega^k_\tau\in\{0,1,2,\dots,T\}$ slots.\footnote{These values are chosen prior to the frame and are sent to the sensors using the beacon signal.} If Sensor $k$ is allocated to time slot $j$ during frame $\tau$, $\omega_{\tau,j}^k=1$; otherwise, $\omega_{\tau,j}^k=0$.

We adopt a block-fading channel model and consider the effect of interference from other sensors that use the same frequency band. In a given time slot, the source knows the channel state information (CSI) of the link connecting itself to its destination, and the link connecting it and the eavesdropper; this assumption is common in the literature, e.g., \cite{krikidis2009relay,6746659}.\footnote{As argued in \cite{6746659}, the CSI is usually estimated through pilots and feedback from the destinations as in \cite{ref32}. CSI estimation without using pilots and feedback may also be implemented as proposed in \cite{ref33}.} Given the availability of CSI at the source in each time slot, the source determines the amount of transmit power needed to mitigate channel outages while maintaining the data undecodable at the eavesdropper (i.e. while maintaining secure communication with the legitimate receiver). In each time slot, the EHS decides whether to transmit its data or remain silent. Remaining silent in a time slot is beneficial for the following reasons: (1) to harvest more energy if the available energy amount does not ensure successful and secure communication, (2) to wait until the channel gains improve over their current levels so that the required power for establishing successful and secure communication becomes lower. We assume that the fading gain\footnote{We refer to the squared absolute value of the complex channel fading coefficient as the \emph{fading or channel gain}.} of a link remains constant during a frame duration, and changes identically and independently from {\it one frame to another}. Note that channels are assumed to be fixed over the whole frame which is reasonable for fixed sensor networks since the coherence time is typically much longer than the frame time.

Assuming additive Gaussian noise and interference, for a given channel realization of the legitimate link, the capacity of the link between the $k$th legitimate EHS and the legitimate destination in time slot $j$ of frame $\tau$ is given by
  \begin{equation}
 \begin{split} \small
 \label{cor}
C^k_L=\log_2\left(1+\frac{P^k_{\tau}\alpha^k_\tau}{\mathcal{N}_{\rm d}}\right)  \,\,\,\,\ \text{[bits/sec/Hz]}
 \end{split}
\end{equation}
where $\mathcal{N}_{\rm d}$ is the noise-plus-interference power at the destination in Watts, $P^k_{\tau}$ is the transmit power (or energy)\footnote{Transmit power and energy are equal in our case as the slot duration is one second, i.e., $T_{\rm s}=1$ seconds as mentioned earlier.} in Watts (joules) of Sensor $k$ at time slot $j$, and $\alpha^k_\tau$ is the squared magnitude of the complex channel fading coefficient between the $k$th sensor and its destination in frame $\tau$.

Similarly, the capacity of the link between the $k$th EHS and the eavesdropper for a given channel realization $\beta^k_\tau$ is
  \begin{equation}
 \begin{split} \small
 \label{cor}
C^k_E=\log_2\left(1+\frac{P^k_{\tau}\beta^k_\tau}{\mathcal{N}_{\rm e}}\right) \,\,\,\,\ \text{[bits/sec/Hz]}
 \end{split}
\end{equation}
where $\beta^k_\tau$ is the squared magnitude of the complex channel fading coefficient between the $k$th sensor and the eavesdropper in frame $\tau$, and $\mathcal{N}_{\rm e}$ is the noise-plus-interference power at the eavesdropper's receiver in Watts. The
secrecy capacity for a given channel fading realization is given by \cite{4035982}
\begin{equation}
\begin{split} \small
\label{romana}
C^k_S = \begin{cases}
 C^k_L-C^k_E &  \alpha^k_\tau/\mathcal{N}_{\rm d} >  \beta^k_\tau/\mathcal{N}_{\rm e}\\
0 &  \alpha^k_\tau/\mathcal{N}_{\rm d} \le  \beta^k_\tau/\mathcal{N}_{\rm e}
\end{cases},
\end{split}
\end{equation}
Let $\mathcal{R}=B/T_{\rm s}/W$ bits/sec/Hz, where $T_{\rm s}=1$ is the slot duration, $B$ is the packet size, and $W$ is the channel bandwidth. As long as the EHS transmits with a rate $\mathcal{R}$ that does not exceed the secrecy capacity, i.e., $\mathcal{R}<C^k_S$, its transmission is secure and successfully decodable at its legitimate destination. Using (\ref{romana}) when $\alpha^k_\tau/\mathcal{N}_{\rm d} >  \beta^k_\tau/\mathcal{N}_{\rm e}$, and the secrecy condition $\mathcal{R}<C^k_S$, we get the following relation
\begin{equation}
\begin{split} \small
 \mathcal{R} <C^k_L-C^k_E&= \log_2\left(1+\frac{P^k_{\tau}\alpha^k_\tau}{\mathcal{N}_{\rm d}}\right)-\log_2\left(1+\frac{P^k_{\tau}\beta^k_\tau}{\mathcal{N}_{\rm e}}\right)
\end{split}
\end{equation}
Letting $\tilde \alpha^k_\tau=\frac{\alpha^k_\tau}{\mathcal{N}_{\rm d}}$ and $\tilde \beta^k_\tau=\frac{\beta^k_\tau}{\mathcal{N}_{\rm e}}$ and after straightforward algebra, we get
\begin{equation}
\begin{split} \small
\label{cor}
P^k_{\tau}&\ge  \frac{2^{\mathcal{R}}-1}{\tilde \alpha^k_\tau-2^\mathcal{R}\tilde \beta^k_\tau}  \,\,\,\,\ \text{[Watts]}
\end{split}
\end{equation}
with $\tilde \alpha^k_\tau-2^\mathcal{R}\tilde \beta^k_\tau>0$. Note that the condition $\tilde \alpha^k_\tau-2^\mathcal{R}\tilde \beta^k_\tau>0$ subsumes $\tilde \alpha^k_\tau > \tilde \beta^k_\tau$. That is, $\tilde \alpha^k_\tau>2^\mathcal{R}\tilde \beta^k_\tau> \tilde \beta^k_\tau$. If $\tilde \alpha^k_\tau-2^\mathcal{R}\tilde \beta^k_\tau\le0$, this implies that the rate exceeds the capacity $C^k_S$; hence, the transmission is not secure. The right-hand side of (\ref{cor}) represents the minimum EHS transmit power needed to ensure that its legitimate link is not in outage (i.e. its data is decodable at the legitimate destination) and to ensure secure communication with the legitimate destination.

One may attempt to optimize the EHS transmit power to satisfy the constraints and to maximize the secrecy throughput. However, due to the fact that the EHS adopts a fixed-rate data transmission scheme, the optimal power is the minimum power that results in successful and secure decoding of the data at the destination. This is because the source (sensor) is  power-constrained and must optimally manage its energy usage. From (\ref{cor}), it is clear that the minimum transmit energy by Sensor $k$ in time slot $j$ of frame $\tau$, denoted by $\mathcal{P}^k_{\tau}$, which is required to ensure that the link between the legitimate source and its destination is secure and not in outage, is $ \mathcal{P}^k_{\tau}=\frac{2^{\mathcal{R}}-1}{\tilde \alpha^k_\tau-2^\mathcal{R}\tilde \beta^k_\tau}$ when $\tilde \alpha^k_\tau>2^\mathcal{R} \tilde \beta^k_\tau$. Hence, one should use this value to achieve the highest throughput as time progresses.

We define a new variable $\gamma^k_{\tau,j}\in\{0,1\}$ which indicates the use of power $\mathcal{P}^k_{\tau}$ or zero power if the source decides to remain idle during the $j$th slot. In particular, if the source uses power $\mathcal{P}^k_{\tau}$, $\gamma^k_{\tau,j}=1$; otherwise $\gamma^k_{\tau,j}=0$.

We denote the energy harvested by Sensor $k$ at the end of the frame (i.e. beginning of the communication and beacon sessions) as $E_{\circ,\tau}^k$. Let $E^k_{\tau,j}$ denote the energy harvested by the $k$th sensor from the ambient sources in time slot $j$ during the $\tau$th frame, and $o^k_{\tau,j}=\gamma^k_{\tau,j} \mathcal{P}^k_{\tau}+\gamma^k_{\tau,j} E^k_{\rm p}$ denote the energy depleted from battery of Sensor $k$ in time slot $j$ due to data transmission and other activities, e.g., processing, measurements, etc. Note that $E^k_{\tau,1}$ can be viewed as the energy harvested at the end of the beacon signal transmission. We assume that at the beginning of frame $\tau$, the battery maintains $\mathcal{B}^k_{\tau-1}$ energy units which is the remaining energy from the previous frame. Note that $\mathcal{B}^k_0$ represents the initial battery energy level of Sensor $k$, i.e., first usage of the battery. In addition, if $\mathcal{B}^k_{\tau-1}+E_{\circ,\tau}^k=0$ or $ \mathcal{B}^k_{\tau-1}\!+\!E_{\circ,\tau}^k\!<\!\tilde E_{\tau}^k$, the sensor will not be able to communicate with the destination (or overhear the beacon) and it remains idle to save its energy; hence, the destination assigns no data slots to that sensor.

The energy stored in the EHS's energy queue at the beginning of the $j$th time slot during the $\tau$th frame, denoted by $Q^{k}_{\tau,j}$, is given by
  \begin{equation}
 \begin{split} \small
 Q^{k}_{\tau,1}&=\mathcal{B}^k_{\tau-1}+E_{\circ,\tau}^k-\delta_\tau^k \tilde E_\tau+E^k_{\tau,1}, \ j=1 \\
 Q^{k}_{\tau,j}&=Q^{k}_{\tau,j-1}- o^k_{\tau,j-1}+ E^k_{\tau,j}, \ \forall j\ge2
 \end{split}
\end{equation}
where $E_{\circ,\tau}^k$ is the energy harvested at the beginning of the frame, $\tilde E_\tau$ is the energy depleted at the beginning of frame $\tau$ for information exchange and beacon decoding, $\delta_\tau^k=1$ if $\mathcal{B}^k_{\tau-1}\!+\!E_{\circ,\tau}^k\!\ge\!\tilde E_{\tau}^k$ and zero otherwise, $o^k_{\tau,j-1}$ represents the energy consumed in time slot $j-1$ of frame $\tau$, and $E^k_{\tau,j}$ is the energy harvested at the beginning of time slot $j$. From the evolution of the energy queue, we note that
 \begin{equation}
 \begin{split} \small
\mathcal{B}^k_{\tau-1}= Q^{k}_{\tau-1,T}-o^k_{\tau-1,T}
 \end{split}
\end{equation}

The causality constraint is given by
  \begin{equation}
 \begin{split} \small
\mathcal{B}^k_{\tau-1}\!+\!E_{\circ,\tau}^k\!+\!\sum_{j=1}^{\ell} E^k_{\tau,j} \!\ge\! \delta_\tau^k \tilde E_\tau\!+\!\sum_{j=1}^\ell o^k_{\tau,j}, \ \forall \ \ell \in\{1,2,\dots,T\}
 \end{split}
\end{equation}
which means that the total energy used up to time slot $\ell$ should not exceed the total energy harvested from the environment up to time slot $\ell$~\cite{reddy,ref8,ref10,6657835}.

 The secrecy throughput of the $k$th EHS during frame $\tau$ is
  \begin{equation}
\small \begin{split} \small \small  \small
     \mu^k_\tau=  \frac{1}{T_f} \sum_{j=1}^T   \gamma^k_{\tau,j} \omega_{\tau,j}^k\mathcal{R} \,\,\,\,\ \text{[bits/sec/Hz]}
     \end{split}
\end{equation}
where $\tilde \mu^k_\tau=\sum_{j=1}^T  \gamma^k_{\tau,j} \omega_{\tau,j}^k$ is the total number of packets transmitted during frame $\tau$ while $\frac{\tilde \mu^k_\tau \mathcal{R}}{T_f}$ is the number of transmitted bits per second per unit frequency. Since the communication between the $k$th EHS and its destination is unsecured if $\tilde \alpha^k_\tau\le \tilde \beta^k_\tau$ or when the transmission rate exceeds $C^k_S$, which happens when $\tilde \alpha^k_\tau\le 2^{\mathcal{R}} \tilde \beta^k_\tau$, the EHS must remain silent when $\tilde \alpha^k_\tau\le 2^{\mathcal{R}} \tilde \beta^k_\tau$, i.e., it should not be assigned any time slots during the current frame; that is, $\omega^k_\tau=\omega_{\tau,j}^k=0, \ \forall j$.
The secure sum-throughput maximization
problem subject to energy causality is formulated as follows:
  \begin{equation}
\small \begin{split} \small \small  \small
\label{pogp}
    & \underset{\substack{{\gamma^k_{\tau,j},\omega_{\tau,j}^k\in\{0, 1\}}}}{\max.}  \,\,\,\,\ \sum_{k=1}^{\mathcal{M}}\mu^k_\tau=\frac{1}{T_f} \sum_{k=1}^{\mathcal{M}} \sum_{j=1}^T   \gamma^k_{\tau,j} \omega_{\tau,j}^k \mathcal{R} \,\,\,\,\ \text{[bits/sec/Hz]}  \ \\&    \hbox {\rm s.t.}   \   \ \mathcal{B}^k_{\tau-1}\!+\!E_{\circ,\tau}^k\!-\! \delta_\tau^k \tilde E_\tau\!+\!\sum_{j=1}^{\ell} E^k_j \!\ge\! \sum_{j=1}^\ell \omega_{\tau,j}^k (\gamma^k_{\tau,j} \mathcal{P}^k_{\tau}+\gamma^k_{\tau,j} E^k_{\rm p}), \forall \ell,k
    \\&  \,\,\,\,\,\,\,\,\,\,\,\ \omega^k_{\tau,j}\!=\!0  \ \hbox{if $\tilde \alpha^k_\tau\!<\!2^\mathcal{R}\tilde \beta^k_\tau $, $ \mathcal{B}^k_{\tau-1}\!+\!E_{\circ,\tau}^k\!=\!0$},\\& \,\,\,\,\,\,\,\,\,\,\,\,\,\,\,\,\,\,\,\,\,\,\,\,\,\,\,\,\,\,\,\,\,\,\,\,\,\,\,\,\ \hbox{or \ $ \mathcal{B}^k_{\tau-1}\!+\!E_{\circ,\tau}^k\!-\!\tilde E_{\tau}^k\!<\!0, \forall j,k$}
     \end{split}
\end{equation}
For given assignments, this problem is a {\it convex integer (binary) program} \cite{boyed} and can be solved efficiently using a standard convex optimization solver such as Matlab's CVX.

An alternative formulation starts with a table of $\mathcal{M}\times T$ elements in which the columns represent the time slots while the rows represent the sensors. We assign only one sensor to each column. i.e., we only have one element with the unity value and the other elements are zeros. Letting $\Gamma_{\tau,j}^k=1$ if Sensor $k$ is allocated to and will access time slot $j$ within frame $\tau$, the optimization problem can be restated as follows
 \begin{equation}
\small \begin{split} \small \small  \small
\label{pogp}
    & \underset{\substack{{\Gamma^k_{\tau,j}\in\{0, 1\}}}}{\max.}  \,\,\,\,\ \frac{ \mathcal{R}}{T_f} \sum_{k=1}^{\mathcal{M}} \sum_{j=1}^T  \Gamma^k_{\tau,j}  \,\,\,\,\ \text{[bits/sec/Hz]}  \ \\&    \hbox {\rm s.t.}   \   \ \frac{\mathcal{B}^k_{\tau-1}\!+\!E_{\circ,\tau}^k\!-\!\delta_\tau^k \tilde E_\tau\!+\!\sum_{j=1}^{\ell} E^k_j}{\mathcal{P}^k_{\tau}+E^k_{\rm p}} \!\ge\! \sum_{j=1}^\ell \Gamma^k_{\tau,j}, \forall \ell,k \
     \\&    \,\,\,\,\,\,\,\,\,\ \sum_{k=1}^\mathcal{M}  \Gamma^k_{\tau,j}\le 1, \forall j
    \\&  \,\,\,\,\,\,\,\,\,\ \Gamma^k_{\tau,j}\!=\!0  \ \hbox{if $\tilde \alpha^k_\tau\!<\!2^\mathcal{R}\tilde \beta^k_\tau $, $ \mathcal{B}^k_{\tau-1}\!+\!E_{\circ,\tau}^k\!=\!0$},\\& \,\,\,\,\,\,\,\,\,\,\,\,\,\,\,\,\,\,\,\,\,\,\,\,\,\,\,\,\,\,\,\,\,\,\,\,\,\,\,\,\ \hbox{or \ $ \mathcal{B}^k_{\tau-1}\!+\!E_{\circ,\tau}^k\!-\!\tilde E_{\tau}^k\!<\!0, \forall j,k$}
     \end{split}
\end{equation}
This problem is a {\it convex integer (binary) program} and can be solved using Matlab's intlinprog or CVX \cite{cvx}. The problem is solved at the destination and the optimal time slot assignments are sent to the sensors inside the beacon signal.
\vspace{-0.2cm}
\section{Simulations and Conclusions}
\vspace{-0.1cm}
\begin{figure}
  \centering
  \includegraphics[width=0.95\columnwidth]{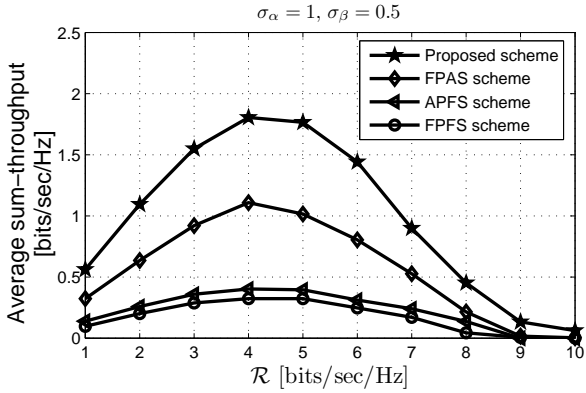}\\
  \caption{Secrecy average sum-throughput versus $\mathcal{R}$ for the considered schemes.}
  \label{fig1}
  \vspace{-0.2cm}%
  \end{figure}
%\vspace{-0.2cm}
In this section, we evaluate the performance of our proposed multiple-access scheme. The simulations are generated using three sensors, i.e., $\mathcal{M}=3$, and noise-plus-interference power levels for the legitimate and eavesdropper links of $\mathcal{N}_{\rm d}=0.1$ and $\mathcal{N}_{\rm e}=1$ milliWatts (mW), respectively. The random channel coefficients are modeled as circularly-symmetric complex Gaussian random variables with zero means and variances $\sigma_\alpha^2$ and $\sigma_\beta^2$ for the legitimate and eavesdropper links, respectively. We assume, without loss of generality, that $\sigma_\alpha=1$ and that the communication between the destination and the sensors lasts for $T_{\rm c}=2$ seconds. The frame is composed of $T=6$ data time slots where a slot duration is normalized to $T_{\rm s}=1$ seconds. Thus, the duration of the frame is $T_f=T_{\rm c}+T=8$ seconds. The energy consumed by the circuits is $E^k_{\rm p}=20$ millijoules (mJ) for Sensor $k$. For our numerical results, we assume that the energy arrival rate is $P_h=10$ mW/second for all sensors and therefore $E^k_1=P_h\times T_{\rm c}=20$ mJ, $E^k_j=P_h \times T_{\rm s}=10$ mJ for all $j\ge 2$, and $E_{\circ,\tau}^k=P_h\times T_{\rm s}=P_h$. The energy needed for the communication among nodes and beacon reception is $\tilde E_{\tau}=0.1$ J for all $\tau$. We assume that all energy queues are initialized at the level $\mathcal{B}^k_{0}=0.11$ J for all $k$. The results are averaged over $\mathcal{F}=1000$ random channel realizations.

 In Fig. \ref{fig1}, we show the average secrecy sum-throughput for our proposed scheme in bits/sec/Hz, which is given by the sum over $k$ and $\tau$ of the multiplication of the total number of transmitted data packets during frame $\tau$, denoted by $\tilde \mu^k_\tau$, and $\mathcal{R}$, i.e., $\mu=\frac{1}{T_f \mathcal{F}}\sum_{\tau=1}^{\mathcal{F}}\sum_{k=1}^\mathcal{M}\tilde \mu_\tau^k \mathcal{R}$ bits/sec/Hz. It is clear that $\tilde \mu^k_\tau$, which is a function of channel outage, decreases monotonically with $\mathcal{R}$ and hence its summation over $k$ and $\tau$. Therefore, the throughput in bits/sec/Hz increases at low $\mathcal{R}$ values, reaches a maximum, and then decreases at high $\mathcal{R}$ values. At $\mathcal{R}=4$ bits/sec/Hz, the throughput gains of our proposed scheme relative to the fixed power with probabilistic (adaptive) time slot allocation (FPAS), fixed power with fixed time slot allocation (FPFS), where each sensor is assigned a set of time slot permanently, and adaptive power with fixed time slot allocation (APFS) schemes are $63\%$, $458\%$, and $349\%$, respectively. We assume that $\sigma_\beta=0.5$. For fixed power schemes, the transmit power is $10$ mW. For fixed slot allocation scheme, we assume that Sensors $1$, $2$, and $3$ are always allocated to data slots $(1,2)$, $(3,4)$ and $(5,6)$, respectively.

\begin{figure}
  \centering
  \includegraphics[width=0.95\columnwidth]{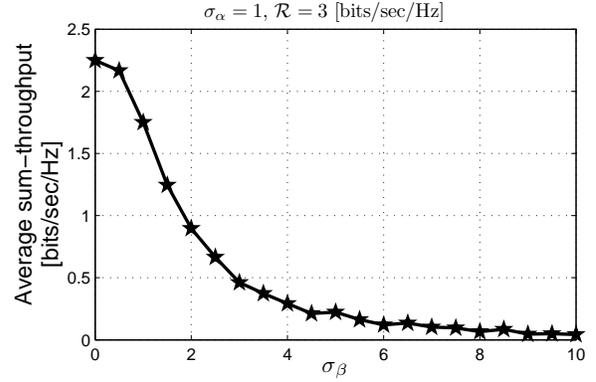}\\
  \caption{Secrecy average sum-throughput versus $\sigma_{\beta}$ for our proposed scheme.}
  \label{fig2}
  %\vspace{-0.2cm}
 %% \vspace{-0.6cm}%
  \end{figure}

In Fig. \ref{fig2}, we plot the average sum-throughput of our proposed scheme versus the standard deviation of the eavesdropper channel $\sigma_\beta$. The case of no eavesdropper is represented by $\sigma_\beta=0$. Fig. \ref{fig2} demonstrates that the presence of an eavesdropper and increasing $\sigma_\beta$ decrease the secrecy sum-throughput. This is because the eavesdropper's capability of decoding the sensors' data increases with $\sigma_\beta$ (since the secrecy capacity decreases and hence the outage probability increases). Figure \ref{fig2} is generated assuming $\mathcal{R}=3$ bits/sec/Hz.
\bibliographystyle{IEEEtran}
\vspace{-0.2cm}%
\bibliography{IEEEabrv,term_bib}
\vspace{-0.2cm}%
\end{document}